# Injection-seeded tunable mid-infrared pulses generated by difference frequency mixing

Yuki Miyamoto,[†*] Hideaki Hara,[†] Takahiko Masuda,[†] Takahiro Hiraki, Noboru Sasao, and Satoshi Uetake

*Research Institute for Interdisciplinary Science, Okayama University, Okayama 700-8530, Japan*

[*]E-mail: miyamo-y@okayama-u.ac.jp

We report on the generation of nanosecond mid-infrared pulses having frequency tunability, a narrow linewidth, and a high pulse energy. These pulses are obtained by frequency mixing between injection-seeded near-infrared pulses in potassium titanyl arsenate crystals. A continuous-wave external cavity laser diode or a Ti:sapphire ring laser is used as a tunable seeding source for the near-infrared pulses. The typical energy of the generated mid-infrared pulses is in the range of 0.4–1 mJ/pulse. The tuning wavelength ranges from 3142 to 4806 nm. A narrow linewidth of 1.4 GHz and good frequency reproducibility of the mid-infrared pulses are confirmed by observing a rovibrational absorption line of gaseous carbon monoxide at 4587 nm.

---

[†]These authors contributed equally to this work.

# 1. Introduction

The control of population and coherence of quantum states by photons is an important technique in many research fields such as quantum chemistry, quantum information, and quantum optics. One such type of control is performed with ultrafast lasers. There are many theoretical and experimental studies on the ultrafast control of atoms, molecules, and semiconductors.[1-11] Because of their short time scale and high peak intensity, it is relatively easy to observe coherent phenomena having time scales limited by the characteristic decoherence time. The drawback of these methods is a relatively small pulse energy of the ultrafast lasers, which makes it difficult to excite a macroscopic number of target atoms or molecules. Experiments in longer time scales are other possible techniques, such as adiabatic passages,[12-16] optical pumping,[17] and saturation spectroscopy[18] using nanosecond pulses or continuous waves. Their high energy enables the transfer of a large population and their narrow linewidth makes the precise control of a single quantum state possible. On the other hand, the observation of coherent phenomena is more difficult in longer time scales owing to decoherence.

Coherent phenomena with high excitation density are an interesting and important topic because their rate is proportional to the excitation density. Recently, it has been claimed that coherent amplification, one of such phenomena, can contribute to solving problems concerning the properties of neutrinos.[19] In these applications, media with both high density and high coherence are desired, for which light sources with both high pulse energy and narrow linewidth are required. Intense nanosecond pulses can fulfill these conditions. Furthermore, frequency tunability is also desired. In previous studies, we excited the vibrational state of hydrogen molecules coherently with two visible lasers and observed the coherent amplification of the two-photon emission from them.[20,21] Although many demonstrations of quantum control have been performed using visible or near-infrared (NIR) lasers, those using mid-infrared (MIR) pulses are not versatile because it is still difficult to prepare MIR lasers with a high pulse energy and a narrow linewidth. For example, such lasers can directly control the vibrational states of molecules. The development of these MIR lasers can pave the way to new methods of quantum control.

For generating tunable nanosecond MIR pulses, frequency mixing techniques are convenient. Frequency tunability and a narrow linewidth can be achieved by using a cavity or a grating. However, it is difficult to generate Fourier-transform-limited pulses using only a cavity or a grating. Injection seeding with a narrow-linewidth laser is another candidate for generating tunable-narrow linewidth pulses. Although it is possible to prepare such MIR pulses by injecting a tunable continuous-wave (cw) MIR laser into optical parametric generation (OPG) systems, MIR lasers with sufficient intensity, linewidth, and tunability are still at the stage of research and development. On the other hand, difference frequency generation (DFG) between injection-seeded NIR pulses can produce narrow-linewidth MIR pulses using relatively simple and inexpensive systems that can be constructed from commonly used devices.

Lithium niobate is often used for generating MIR pulses.[22-28] However, its absorption increases at wavelengths longer than 4 μm.[29] Silver thiogallate ($AgGaS_2$) and silver gallium selenide ($AgGaSe_2$) are widely used to generate longer wavelength MIR photons.[30-32] However, their low threshold for laser-induced damage makes high-power generation difficult.[29] Recently, high-power MIR generation has been studied intensely using other crystals such as zinc germanium phosphide ($ZnGeP_2$).[33-38] These crystals are still expensive and not commonly available. Furthermore, $ZnGeP_2$ needs a pumping source of approximately 2 μm. In order to achieve a high power, a narrow linewidth, and a wide tuning range with commonly used devices, we perform DFG in potassium titanyl arsenate ($KTiOAsO_4$, KTA) as a nonlinear medium. Kung developed an injection-seeded MIR nanosecond pulsed laser system by DFG in KTA.[39] The system was capable of generating MIR pulses from 3.0 to 5.3 μm using two Nd:YAG pumping lasers. Kung reported that the linewidth was within 1 cm$^{-1}$ (30 GHz). We have also constructed an MIR laser system that generated nanosecond pulses at 4587 nm via DFG between seeded NIR pulses in KTA crystals. In our previous study, we used a prototypical system to trigger a two-photon emission.[21] In the present study, we refine it and achieve a higher output intensity and a wider tuning range. Its properties, such as linewidth, phase-matching conditions, and output energies, are investigated in detail. We demonstrate wide tuning range by actually changing the wavelength of the MIR pulses from 3142 to 4806 nm. This region covers vibrational energies of many molecules, for example, CO, $CO_2$, and various alkanes. The advantages of our laser over that described in Ref. 39 are a narrower linewidth of MIR pulses and a simpler system with a single pumping source. The absorption spectrum of carbon monoxide (CO) is measured to estimate the linewidth and frequency reproducibility of the present system.

## 2. Experimental methods

Figure 1 shows a schematic of the MIR laser system. The MIR pulses are prepared in the three steps described below. In the first step, a MgO-doped periodically poled lithium niobate (PPMgLN) crystal (Covesion MSFG1-20) of 20 mm length and 0.5×0.5 mm$^2$ aperture size is pumped by the second harmonics (λ = 532.3 nm) of a seeded pulsed Nd:YAG laser. The pulse duration of the second harmonics is 4 ns and its linewidth (full width at half maximum, FWHM) is estimated to be less than 200 MHz using a Fabry-Perot interferometer. The typical pumping energy is 130 μJ/pulse. The temperature of PPMgLN is controlled by a temperature controller. PPMgLN is seeded using a cw beam from an external cavity laser diode (ECLD) system or a Ti:sapphire ring laser (Coherent 899-29). The pumped PPMgLN emits signal ($\nu_{sig}$) and idler ($\nu_{idler}$) pulses by OPG. In the present system, the wavelength of $\nu_{sig}$ is tuned spontaneously to that of the seeding beam under appropriate seeding conditions so that it is possible to control the signal wavelength by adjusting the seeding wavelength. The wavelengths of $\nu_{sig}$ and $\nu_{idler}$ are 795.2–871.5 and 1367–1609 nm in the present experiment, respectively. The linewidths of the generated pulses are narrowed by seeding. The ECLD system consists of a laser diode (Eagleyard Photonics EYP-RWE-0870-06010-0750-SOT01-0000), an external cavity, and a tapered amplifier (m2k-laser m2k-TA-0870-1500). The typical seeding power and linewidth of the ECLD system are 1 W and 2 to 3 MHz, respectively. The tuning range of the ECLD system is 865–880 nm, which is limited by the bandwidth of the tapered amplifier. The power and tuning range of the Ti:sapphire laser are about 300 mW and 650–900 nm, respectively. Using the ECLD as the seeder makes the system smaller, simpler, and more inexpensive, while using the Ti:sapphire laser makes the frequency tuning range wider. The pumping pulses and seeding beam are collimated by lens pairs to a diameter of 0.5 mm.

In the second step, the generated $\nu_{sig}$ pulses are injected into lithium triborate (LBO) crystals (Tecrys, $\theta = 90°$ and $\varphi = 0°$), which are pumped by the 532.3 nm pulses from an identical Nd:YAG laser to that in the first step (typically 50 mJ/pulse). Because of the optical parametric amplification (OPA), the intensities of $\nu_{sig}$ and $\nu_{idler}$ increase. We use four collinearly aligned LBO crystals of 4×4 mm$^2$ aperture size and 120 mm total length. The temperature of the crystals is controlled in an oven. Type-I noncritical phase matching allows a relatively large gain length.

In the third step, the amplified $\nu_{idler}$ pulses and 1065 nm pulses (50 mJ/pulse, 2 mm in diameter) from the identical Nd:YAG laser are collinearly introduced into KTA crystals (Crystech, $\theta = 41°$ and $\varphi = 0°$) at room temperature. MIR pulses (3142 to 4806 nm in the present experiment) are generated by the DFG process between $\nu_{idler}$ pulses and 1065 nm pulses in the KTA crystals. We use five or six KTA crystals of $3\times3$ or 4×4 mm$^2$ aperture size. The total length is 50 or 60 mm. Each crystal is mounted on a separate holder and its angle can be independently controlled so as to satisfy the type-II critical phase-matching condition. Because of the birefringence of the KTA crystals, the $\nu_{idler}$ beam deviates from the other two beams with a walk-off angle during their propagation in the crystals. The walk-off angle of extraordinary waves in uniaxial crystals is represented as $-\arctan[(n_o/n_e)^2 \tan\theta] + \theta$, where $n_o$ ($n_e$) is the refractive index for an ordinary (extraordinary) wave and $\theta$ is the angle between the Z-axis and the wave vector.[40] Although KTA crystals are biaxial, this equation can be applied to the current case of $\varphi = 0°$ by assuming $n_o = n_X$ and $n_e = n_Z$. When the $\nu_{idler}$ beam of 1400 nm is injected perpendicularly to the input facet, $\theta$ is equal to the cut angle of 41°, and the walk-off angle of the $\nu_{idler}$ beam is 22 mrad. The refractive indices are taken from Ref. 29. The estimated walk-off angle indicates that the $\nu_{idler}$ beam deviates 1.3 mm from the other beam after 60 mm propagation. This deviation is not negligible for the beam diameters used. Therefore, the crystals are aligned so that each y-axis direction is opposite to that of adjacent crystals to cancel walk-off.

To evaluate the frequency properties of the MIR pulses in detail, the absorption spectrum of the CO R(9) rovibrational transition at 2179.8 cm$^{-1}$ (4587.6 nm) is observed.[41] The pulse energy is reduced by a factor of about $10^{-6}$ using neutral density filters to avoid saturation broadening. The cell length is 150 mm and the CO pressure is 13 Pa. Transmitted MIR pulses are detected using a mercury-cadmium-telluride detector (Daylight Solutions HPC-2TE-100). The output voltage of the detector is monitored using an oscilloscope and sent to a computer. The MIR frequency is scanned by changing the frequency of the Ti:sapphire laser. The scan width is set to 10 GHz in the present experiment. The temperatures and angles of the crystals are constant during the scan. A single scan takes about 30 s. Frequency axes of the spectra are calibrated using the scan drive signal from the control unit of the laser. The interval between data points of the observed spectra is about 30 MHz. The data points are binned at 100 MHz intervals for analysis.

## 3. Results and discussion

Figure 2 shows observed (blue squares) and calculated (solid lines) quasi-phase-matching conditions for the first step (OPG in PPMgLN). The typical output energy of $\nu_{sig}$ pulses from PPMgLN is 20 μJ/pulse and does not depend strongly on their wavelengths in the present experiment (795.2–871.5 nm). It is found that the seeding power should be higher than 200 mW to ensure the seeding effects. When it is lower than 200 mW, the linewidth of the generated MIR becomes broad and the frequency cannot be controlled. It is known that the seeding power should be larger than a threshold in order to overcome spontaneous emission and show an efficient seeding effect.[42] Similar phenomena occur when the pumping energy becomes high; hence, we keep it below ~150 µJ/pulse. The calculation for phase-matching conditions is based on the work of Li et al.[43] The temperature-dependent Sellmeier equation for congruent lithium niobate is represented as

$$n_e^2 = a_1 + b_1 f + \frac{a_2 + b_2 f}{\lambda^2 - (a_3 + b_3 f)^2} + \frac{a_4 + b_4 f}{\lambda^2 - a_5^2} - a_6 \lambda^2, \qquad (1)$$

where $a_i$ and $b_i$ are given in Ref. 44, λ is the wavelength in µm, and $f$ is a function of the crystal temperature $T$ (°C) and is given by

$$f = \frac{T - 24.5}{T + 570.82} . \qquad (2)$$

Similarly to Ref. 43, $a_1$ is used as a free parameter and determined to reproduce experimental results. The resultant $a_1$ is 5.64. The shift from the value for the congruent lithium niobate (5.36) is slightly larger than that in Ref. 43 (5.53). The experimental and calculated results are consistent (Fig. 2).

Figure 3 shows the observed (blue squares) and calculated (solid lines) phase-matching conditions for the second step (OPA in LBO). The frequency range of the observed data is 1367–1610 nm, which corresponds to the MIR range of 4806–3142 nm in the DFG step. The typical output energies of $v_{idler}$ pulses are 3 to 5 mJ/pulse at approximately 1400 nm and 1 mJ/pulse at approximately 1500 nm. When LBO crystals with a total length of 80 mm were used, the $v_{idler}$ energy was below 1 mJ/pulse at approximately 1400 nm; therefore, we adopted crystals of 120 mm total length. The calculation for the phase-matching condition is basically the same as that reported by Lin et al.[45] The broken line shows the calculated condition shifted by 3.5 °C. The shifted line and experimental data (blue squares) are consistent. This deviation of 3.5 °C between the calculation and the experiment may be due to inhomogeneity of the crystal temperature because the LBO oven is large. As a result, the effective temperature of LBO deviates from the measured one. Furthermore, the effective temperature depends on the environment. The variation in the experimental data of about three degrees at approximately 1.4 µm in Fig. 3 is attributed to this effect because these data were taken at separate periods.

In the third step (DFG in KTA), the energy of the generated MIR pulses is typically 0.4 mJ/pulse at approximately 4500 nm and 1 mJ/pulse at approximately 3500 nm (Fig. 4). The MIR energy is related almost linearly to the total crystal length up to 40 mm. The ramp rate is approximately 0.01 mJ/mm at approximately 4500 nm. Although it shows saturation with crystal lengths larger than 40 mm, the output energy slightly increases. Therefore, we use five or six KTA crystals equivalent to a total length of 50 or 60 mm. The standard deviation of the output pulse energy is approximately 5% of the average energy. The output energy is almost linear relative to the pumping energy. A higher output pulse energy can be achieved by increasing the pump laser intensity, because the laser-induced damage threshold of KTA crystals (> 1.2 GW/cm$^2$, 8 ns, 20 Hz) is higher than the present intensity of 1065 nm pulses of less than 1 GW/cm$^2$.[29] It is worth mentioning that improving the spatial modes of two NIR pulses also increases the output energy. The type-II phase-matching condition of the DFG is calculated to be 41 to 42° between 3142 and 4806 nm using the Sellmeier equation reported in Ref. 29. A fine angle adjustment of 1° or less around the cutting angle of KTA ($\theta$ = 41°) can fulfill the phase-matching condition. The propagation of the MIR beam profiles measured by a laboratory-build slit scan profiler indicates that $M^2$ for the MIR pulses is approximately 3.

In the present experiment, the MIR frequency is changed from 3142 to 4806 nm to demonstrate the tunability of the system (Fig. 4). The wavelength of the generated MIR pulses is determined from the wavelengths of the Nd:YAG laser and the seeding laser (ECLD or Ti:sapphire laser) monitored using wavemeters (Highfinesse WS-6 and WS-7). Their absolute accuracy is 600 MHz or higher. The wavelength of the Nd:YAG laser is 1065 nm. The frequency tunability of the MIR pulses seeded using the Ti:sapphire laser was examined with a monochromator (Princeton Instruments Acton SP2300) whose resolution is higher than 1 nm. Figure 4 shows examples of normalized spectra of the MIR pulses. Observed peak wavelengths are consistent with the values calculated from the Ti:sapphire laser wavelength within ~1 nm. The accuracy of the MIR wavelength is discussed below using CO gas spectra. The FWHM of each spectrum is about 1 nm, which is limited by the resolution of the monochromator. As described above, the pulse energies at 4429 and 4586 nm are both approximately 0.4 mJ/pulse, while that at 4805 nm is one order of magnitude smaller (Fig. 4). This energy degradation at long wavelengths can be attributed to the absorption by the KTA crystals. The linear absorption coefficient of KTA crystals was reported to be 0.2 cm$^{-1}$ at 4000 nm and 1.0 cm$^{-1}$ at 5000 nm.[29] Therefore, it is difficult to generate intense pulses at wavelengths longer than 4800 nm with the current system. On the other hand, the phase-matching conditions (Figs. 2 and 3) indicate that MIR pulses of 2600 nm can be generated by the current system without modifications when the Ti:sapphire laser is used as the seeder. It is worth mentioning that a continuous tuning range of 10 GHz is achieved, as described in the previous section.

To estimate the frequency accuracy, linewidth, and reproducibility of the MIR pulses, the MIR absorption spectroscopy of CO gas is performed using the MIR pulses from the present system. Figure 5 shows the spectrum of the CO R(9) line at 2179.77 cm$^{-1}$ (4587.64 nm) averaged over eight traces. Its peak position and FWHM are estimated to be 2179.76 cm$^{-1}$ (4587.66 nm) and 1.4 GHz, respectively, by fitting with a Gaussian function. The peak position agrees with the reported transition frequency within the accuracy of the wavemeter (600 MHz), which measures the frequency of the seeding laser ($v_{sig}$). Furthermore, the peak positions of all scans agree within 300 MHz, which indicates good frequency reproducibility of the system. The Doppler linewidth is calculated to be 150 MHz at room temperature; thus, the observed linewidth mainly originates from the laser linewidth. The pressure broadening and shift are negligibly small under the current conditions. Although the reproducibility of the linewidth and frequency is measured only at 4587.66 nm, those for other wavelengths are of the same order when injection seeding works well. The linewidth of 1.4 GHz is one order of magnitude narrower than that in Ref. 39. The time duration of the pulses is measured to be 2 ns. Assuming a Gaussian profile, its Fourier transform limit is calculated to be 200 MHz (FWHM), which is much smaller than the observed spectral width. One of the possible reasons is that the high gain of PPMgLN may cause the undesired generation of unseeded frequency components. Further narrowing may be achieved by using an increased seeding power, a cavity-based PPMgLN oscillator, and better spatial mode matching between the seeding beam and the pumping beam. It is also worth mentioning that the linewidth depends on the alignment of the crystals. The misalignment of the crystals may enhance the undesired components because the phase-matching conditions are not very strict.

## 4. Conclusions

We constructed a tunable narrow-linewidth MIR laser with a high output energy by the injection-seeding DFG technique. MIR pulses from 3142 to 4806 nm were actually generated. Typical emitting energies were 0.4 mJ/pulse at approximately 4500 nm and 1.0 mJ/pulse at approximately 3500 nm. It was difficult to generate high-energy pulses of wavelengths longer than 4800 nm because of the absorption of KTA crystals. A higher output energy can probably be achieved by increasing the pumping intensity. The linewidth was estimated to be 1.4 GHz from the MIR absorption spectra of the CO R(9) line at 4587.66 nm. The frequency reproducibility was also estimated to be within 300 MHz.

**Acknowledgments**

This research was supported by a Grant-in-Aid for Scientific Research on Innovative Areas "Extreme quantum world opened up by atoms" (21104002), a Grant-in-Aid for Scientific

Research A (15H02093), a Grant-in-Aid for Young Scientists B (15K17651), and a Grant-in-Aid for Research Activity start-up (26887026) from the Ministry of Education, Culture, Sports, Science, and Technology.

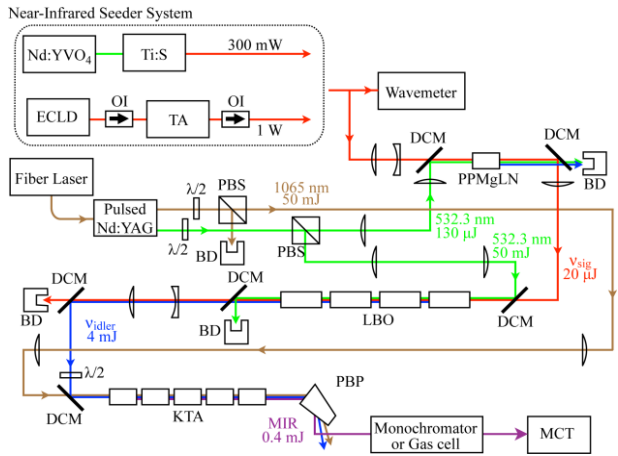

**Fig. 1.** Schematic of the laser system. The wavelength, energy, and power in the figure are typical values. λ/2: half-wave plate, BD: beam dumper, DCM: dichroic mirror, ECLD: external cavity laser diode, KTA: potassium titanyl arsenate, LBO: lithium triborate, MCT: mercury cadmium telluride photodetector, Nd:YVO$_4$: Nd:YVO$_4$ laser (Coherent, Verdi V10), OI: opical isolator, PBP: Pellin-Broca prism, PBS: polarization beam splitter, PPMgLN: MgO-doped periodically poled lithium niobate, TA: tapered amplifier, Ti:S: Ti:sappire ring laser.

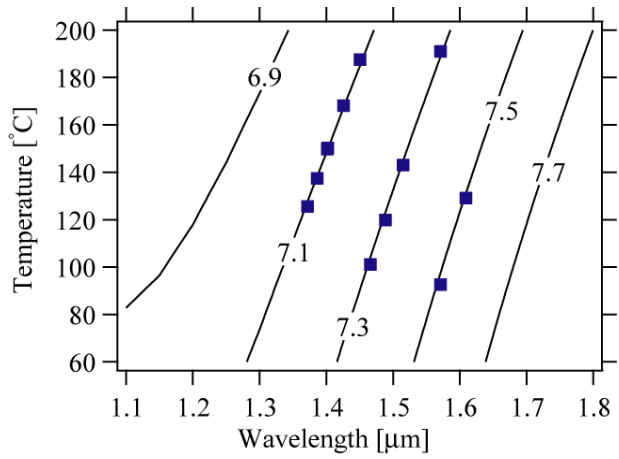

**Fig. 2.** Quasi-phase-matching conditions of PPMgLN OPG pumped by 532.3nm pulses. The vertical axis shows the crystal temperature and the horizontal axis shows the wavelength of the idler. Numbers in the panel show periods of the crystal in μm. Blue squares show experimental data.

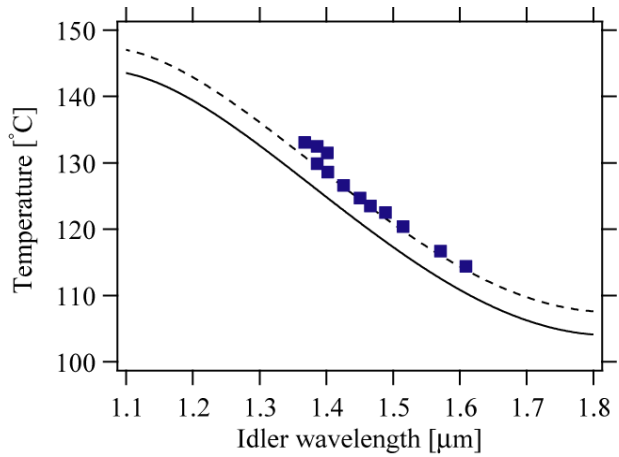

**Fig. 3.** Phase-matching condition of LBO OPA pumped by 532.3 nm pulses. The vertical axis shows the crystal temperature and the horizontal axis shows the wavelength of the idler. The solid line shows the calculated condition. The broken line shows the calculated condition shifted by 3.5 °C. Blue squares show experimental data.

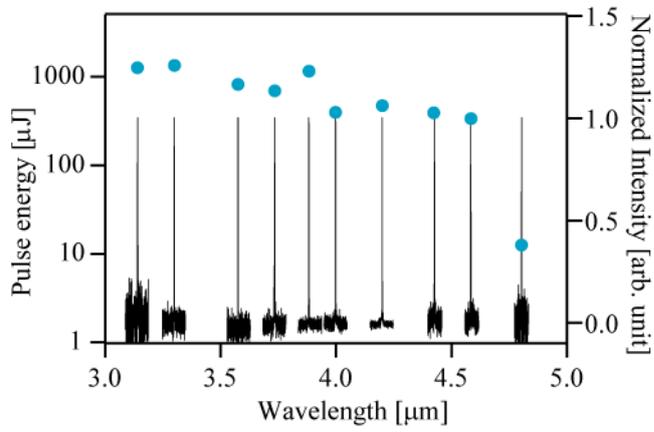

**Fig. 4.** Spectra of the MIR pulses observed with a monochromator (right axis). Each spectrum is normalized by its maximum. Blue closed circles show the typical pulse energies of the MIR pulses at various wavelengths (left axis).

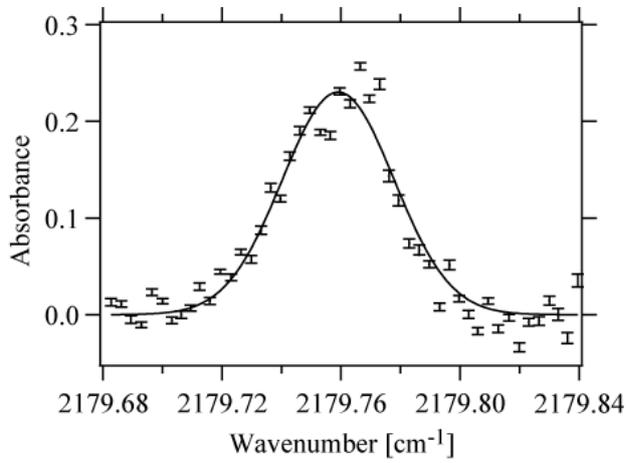

**Fig. 5.** Observed absorption line R(9) of CO gas. The fitted Gaussian curve is also shown. The horizontal axis is converted from the measured wavelength of the seeding laser.